\documentclass[conference]{IEEEtran}
\usepackage{graphicx}
\usepackage{cite}
\begin{document}

\title{IceCube : Toward a km$^3$ Neutrino Telescope}

\author{\authorblockN{Paolo Desiati}
\authorblockA{on behalf of the IceCube Collaboration${^{(\dag)}}$\\
University of Wisconsin - Madison\\
${(\dag)}$ http://icecube.wisc.edu\\
Email: desiati@icecube.wisc.edu}}

%\specialpapernotice{(Invited Paper)}

\maketitle

\begin{abstract}
Since the end of the 2005-2006 austral summer, the IceCube detector consists of an array
of 9 strings, deployed between 1450 m and 2450 m of depth and containing 540 digital
optical sensors and 16 IceTop surface stations with 64 sensors. With the integrated
AMANDA-II experiment this is the world's largest neutrino telescope in operation. The
construction status of IceCube is presented along with an update on physics performance
study for the detection of high energy neutrinos. The potential of the full km$^3$-scale
telescope in the search for astrophysical sources is also addressed.
\end{abstract}


% for peerreview papers, inserts a page break and creates the second title.
% Will be ignored for other modes.
\IEEEpeerreviewmaketitle

\section{Introduction}
\label{s:intro}
% no \PARstart

\begin{figure}[h]
  \centering
  \includegraphics[width=2.5in]{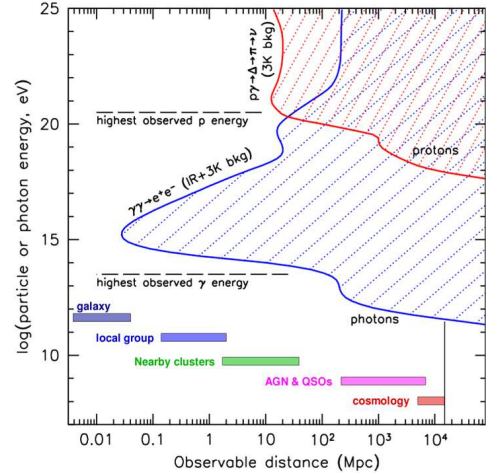}
  \caption{Cosmic Messengers energy vs the distance within which they are detectable. Photons are affected by
	absorption in the infrared and cosmic microwave background : the blue shaded area represents, for any given
	energy, the distance range a photon cannot reach. Protons, besides undergoing deflection due to magnetic fields,
	are also affected by absorption in the cosmic microwave background: the red shaded area gives the distance range a
	proton cannot reach at a given energy.}
  \label{fig:cosmes}
\end{figure}

The first generation neutrino telescopes, such as NT200 at Lake Baikal \cite{baikal} and AMANDA
\cite{amanda}, are the proof of concept for the detection of high energy neutrinos using the
Cherenkov light emitted by the charged leptons produced in charged-current interactions.
% changed
%Neutrinos can propagate through the Universe undisturbed and without being absorbed and deflected by
%magnetic fields, therefore they they point back to their sources. This is the basis of neutrino astronomy.
Neutrino astronomy relies on the fact that these particles can propagate
through the Universe undisturbed and without being deflected by magnetic fields, therefore they point back
to their sources. Gamma ray astronomy is based on the same principle, but the
observable distance, at TeV energy ranges, is limited to a few tens of Megaparsec due to absorption by
pair production in the infrared and cosmic microwave background (see Fig. \ref{fig:cosmes}).

Protons can propagate deeper distances than gamma rays at a given energy, but they are deflected by the
extragalactic and interstellar magnetic fields and, therefore, they cannot provide any pointing
information. Only at energies above $10^{19}$ eV, deflection is so small that proton astronomy becomes
possible. Unfortunately at those high energies proton absorption in the cosmic microwave background (GZK
cutoff \cite{gzk}) limits the range to a few tens of Megaparsec (see Fig. \ref{fig:cosmes}).

Nevertheless the detection of photons and protons with energy in excess of $10^{13}$ eV and $10^{20}$ eV,
respectively, has triggered a profound theoretical dispute on the understanding of the mechanisms
that produce such high energies in the Universe. Correlation of high energy gamma ray emission with
the density of molecular gas in the center of the Galaxy, as detected by H.E.S.S. \cite{hess} for
instance, indicates that the photons could be produced by cosmic rays accelerated in sites such as the
supernova remnant Sgr A East or the black hole Sgr A*. The decay of mesons produced by proton interaction
with matter and radiation, would produce high energy neutrinos as well as gamma rays. Therefore neutrino
detection from the same gamma ray sources would be the definitive evidence for hadronic acceleration
mechanisms in the high energy cosmic ray sources. Similar arguments could be valid for extragalactic sources
of gamma rays, even if promising sources of high energy extragalactic neutrinos could also be those with strong
gamma ray absorption \cite{bere}.

Unfortunately neutrinos are very difficult to detect due to their small interaction cross section, and therefore, large
target volumes are necessary for their detection and identification.
AMANDA has proven the feasibility of high energy neutrino astronomy and has set stringent limits on extraterrestrial
neutrinos \cite{panic05, hill}. Nevertheless a km$^3$-scale detector is most probably the requirement for detecting expected
neutrino fluxes, such as the ones from cosmic ray interactions on the microwave photons (GZK neutrinos), and for
reaching the detection level of high energy neutrinos associated to the measured cosmic rays at 10$^{18}$ eV in
the hypothesis of optically this extra-galactic sources (Waxman-Bahcall bound \cite{wb}).

% changed
The mass composition of cosmic rays is expected to become heavier at the so-called knee of the cosmic
ray energy spectrum, at about 10$^{15}$ eV. The measurement of cosmic ray energy and mass composition with good
resolution can be done by detecting the correlation between the electron component at
the surface and penetrating muon content (see Sec \ref{ss:cr}). The South Pole location, at an altitude of about
2,800 meters a.s.l., is also characterized by reduced fluctuations in atmospheric shower size, providing better
correlation to primary cosmic ray properties.
%Indirect detection of cosmic rays beyond the knee represents a direct probe of the hypothesized transition between
%galactic and extragalactic component. The detailed measurement of mass composition for mass groups, in this energy range
%would reveal the nature of this transition. A measurement of primary cosmic ray energy and mass determination with
%good resolution are essential and 
%this measurement and IceTop, the surface counterpart of IceCube, is designed to measure composition up to $\sim 10^{18}$ eV
%with good resolution.

In this paper we describe the present construction status of the IceCube neutrino telescope and of the integrated surface
array IceTop, as well as their verification and commissioning in view of the incoming physics results.

\section{IceCube: design and construction}
\label{s:design}

The analog technology used in AMANDA, where the signals recorded by the photomultiplier tubes (PMT) are
propagated through the cables to the surface data acquisition, results to be unsuitable for a detector with
a large number of sensors. A km$^3$ array requires semi-autonomous self-calibrating optical sensors. The digital
transmission of PMT signals guarantees no data loss and allows higher dynamic range \cite{performance}.
A Digital Optical Module (DOM) is the basic detection component. It hosts a 10-inch
Hamamatsu photomultiplier tube and its own data acquisition circuitry, making it an autonomous data collection
unit. The PMT and the data acquisition electronics are protected by a pressure-resistant glass sphere.
A custom-made Analog Transient Waveform Digitizer (ATWD), samples the PMT pulse in three different gain regimes
($\times$1/4, $\times$2 and $\times$16) at a rate of 300 MHz and with a depth of 128 bins (i.e. 425 ns).
The same pulse is also digitized
by a 40 MHz fast Analog-to-Digital Converter that records 255 samples over 6.4 $\mu$s. The linear dynamic range
of a DOM is 400 photo-electrons (p.e.) in 15 ns during recording phase of ATWD. The ATWD samples the pulse from
any given DOM only when it has a local trigger with a neighbouring DOM. Each DOM also provides a time stamp
synchronized to the surface GPS. The DOMs are also equipped with a flasher board which contains 12 LEDs that
can be used to produce light pulses for calibration purposes (see Sec \ref{s:calib}).

The waveforms digitized by each IceCube string and IceTop station are sent to the surface data acquisition system, where
data are time sorted. An event builder forms an event after a simple majority trigger is satisfied for IceCube and IceTop,
respectively. An online filtering system at the South Pole selects only a fraction of
the events to be transferred to the Northern Hemisphere through satellite for physics analyses. 

The present array consists of 9 IceCube strings, deployed between 1450 m and 2450 m depth below the South Pole ice surface,
with a total of 540 DOMs (60 per string). And of 32 surface tanks with 2 DOMs, each looking into transparent ice. The whole
array will consist of 80 strings, for a total of 4,800 DOMs and 160 surface tanks, for a total of 320 DOMs.

The construction relies on hole drilling in the ice using an Enhanced Hot Water Drill (EHWD). The drill system
consists of numerous pump and heating devices, hoses, a drill tower and a complex control. The EHWD is designed
to drill holes to a depth of 2500 m in less than 40 hours, excluding the time needed for rigging. This is approximately
3 times shorter than AMANDA drilling system \cite{karle}.

\section{Calibration}
\label{s:calib}

The detector calibration is one of the major efforts aimed at characterizing its response and to reduce
systematic uncertainties at the physics analysis level. Each PMT is tested in order to characterize its response
and to measure the voltage yielding a specific gain, currently at $10^7$ in the operating detector. The gain
measurement accounts for uncorrelated noise in the optical module due to thermal background in the photocathode
and to radioactive decay of isotopes contaminating the glass pressure sphere. The dark noise rate is approximately
700 Hz when DOMs are in ice.

Each DOM has a free running 20 MHz oscillator which is synchronized to the surface master GPS clock every 3 seconds.
A pulse with known characteristics is sent from the surface to each DOM which in turn, after synchronizing its
local oscillator, sends back an identical
pulse after a known time delay \cite{performance}. This procedure has a resolution of less than 2 ns.

The LEDs on the flasher boards are used to measure the photo-electron transit time in the PMT for the reception of
large light pulses between neighboring DOMs. This delay time is given by light travel time from the emitter to
receiver, by light scattering in the ice and by the electronics signal processing. The RMS of this delay is also less
than 2 ns.

Waveform sampling amplitude and time binning calibration is periodically performed in each DOM and used to extract
the number of detected photo-electrons with an uncertainty of $\sim$ 10\%.

IceTop tanks are calibrated using low energy cosmic muons which deposit $\sim$ 190 MeV, producing a signal of
about 240 p.e., and which provide a characteristic peak used to calibrate the surface array.

Higher level calibrations are meant to correlate the number of detected photo-electrons to the energy of physics
events that trigger the detector. A complete detector response simulation is necessary for this and the ice optical
properties are one of the fundamental ingredients. These properties have been measured in the past using AMANDA in-situ
calibration lasers \cite{opto} and recently using a high precision dust logger \cite{logger} that measures
the dust concentration in the ice as a function of depth. The concentration of dust can be used to measure the ice
optical properties.

\section{Physics Analyses}
\label{s:phys}

\subsection{Backgrounds}
\label{ss:back}

IceCube is triggered mainly by the intense rate of muons generated by the impact
of primary cosmic rays in the atmosphere and by a tiny proportion (i.e. five orders of magnitude smaller) of
atmospheric neutrino-induced events. Atmospheric neutrinos represent the irreducible background for
extraterrestrial high energy neutrino searches and, therefore, it is important to understand the theoretical
uncertainties derived from predictions, as well as experimental systematics, since they might affect the
estimation of the sensitivity for cosmic neutrinos.

\begin{figure}[t]
  \centering
  \includegraphics[width=2.5in]{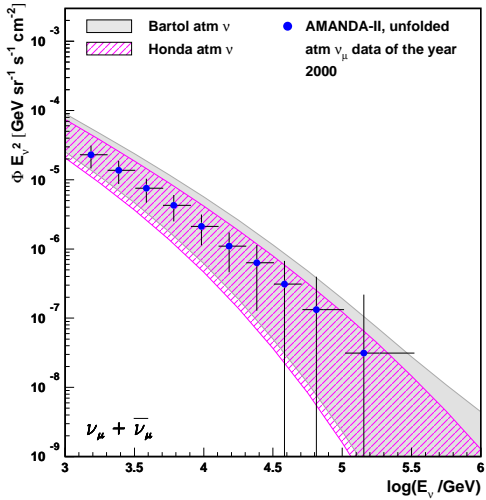}
  \caption{Preliminary unfolded energy spectrum of atmospheric muon neutrinos measured by AMANDA-II using data
		collected between in year 2000. Error bars include statistics and the correlation
		between neighboring energy bins \cite{kirsten}. The theoretical atmospheric neutrino prediction
		shown are from
		\cite{barr} (Bartol) and from \cite{honda} (Honda). Bands are determined by the horizontal
		and verical flux.}
  \label{fig:atmnu}
\end{figure}

Since we use the Earth to discriminate neutrinos from down-ward cosmic muons, only up-going muon tracks are selected 
to reduce the contamination of down-ward cosmic muon events. These up-going events are still contaminated by less
than 1\% wrongly reconstructed cosmic muons, which is too large. Quality cuts are designed to reduce this contamination.
About a 10$^6$ background rejection power needs to be achieved, while mantaining the highest possible neutrino selection efficiency.

Fig. \ref{fig:atmnu} shows the preliminary energy spectrum of atmospheric neutrinos measured with AMANDA-II using
data recorded during in year 2000. A regularized unfolding technique has been used and bin-to-bin correlation
accounted in the error bars \cite{kirsten}. The unfolded energy spectrum is consistent with
present-day uncertainties from cosmic ray spectrum and composition, hadronic interaction models and detector
response modelling.

\vskip 0.4cm
\begin{figure}[h]
  \centering
  \includegraphics[width=2.4in,angle=-90]{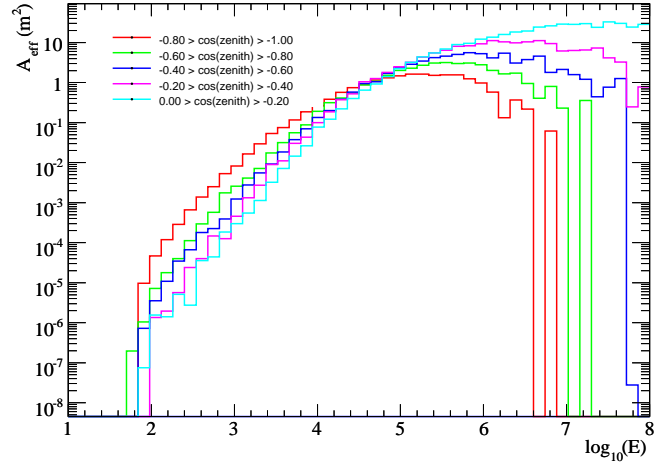}
  \caption{Preliminary neutrino Effective Area, as a function of neutrino energy (in GeV), for selecting
		atmospheric neutrinos using data collected in the year 2006
		by the 9 strings of IceCube. The different histograms corresponds to different angular ranges.}
  \label{fig:ic9}
\end{figure}

A preliminary study of atmospheric neutrino selection has been done with the first 9 strings of IceCube using data
collected and filtered online in 2006, demonstrating the stability during data taking, the reliability of the
online event filtering and the mature stage of analysis tools. The final sample has been selected to
have high quality events and no more than 5\% cosmic muon background contamination, according to simulation. 

\begin{figure}[t]
  \centering
  \includegraphics[width=3.0in]{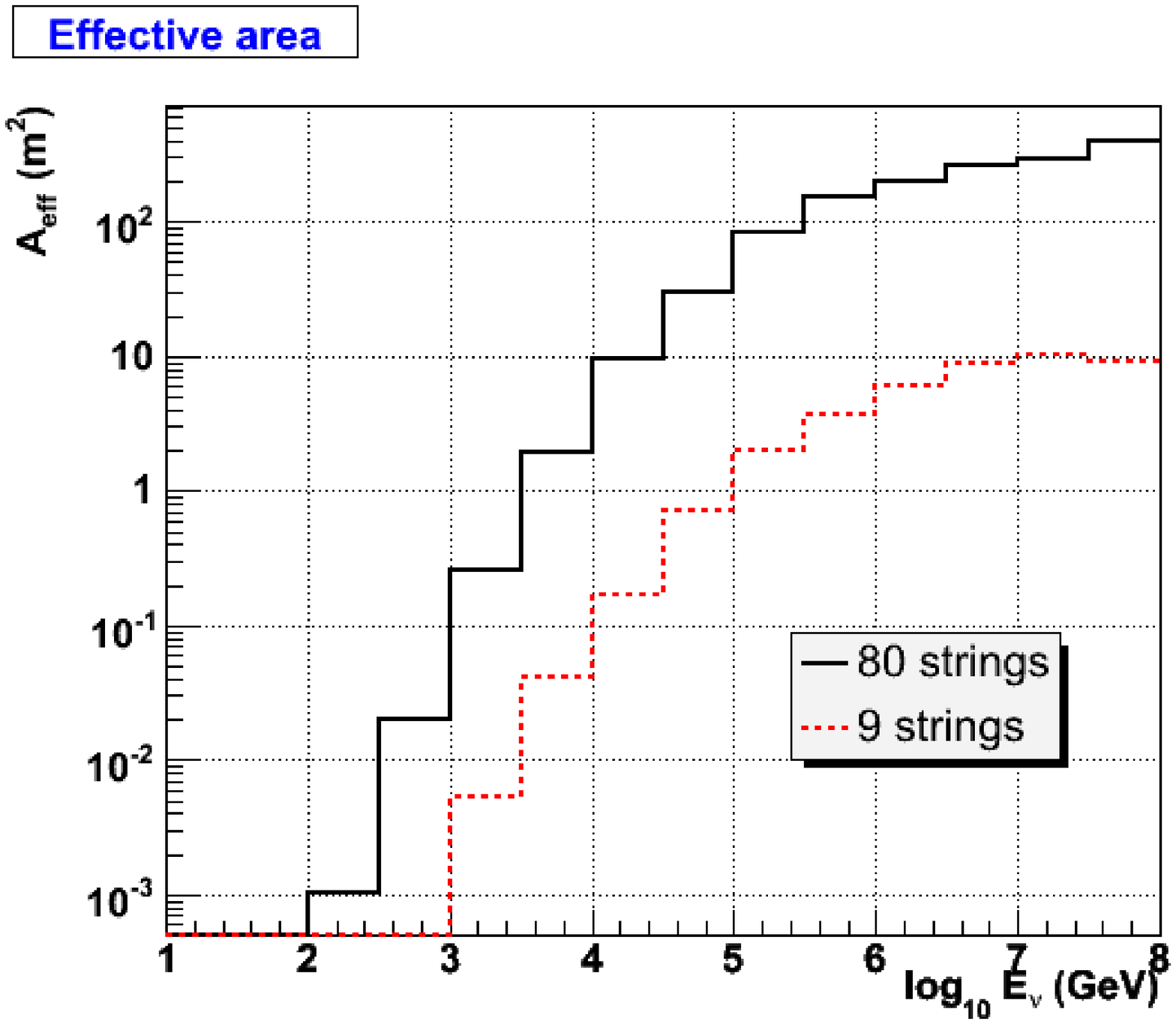}
  \includegraphics[width=3.0in]{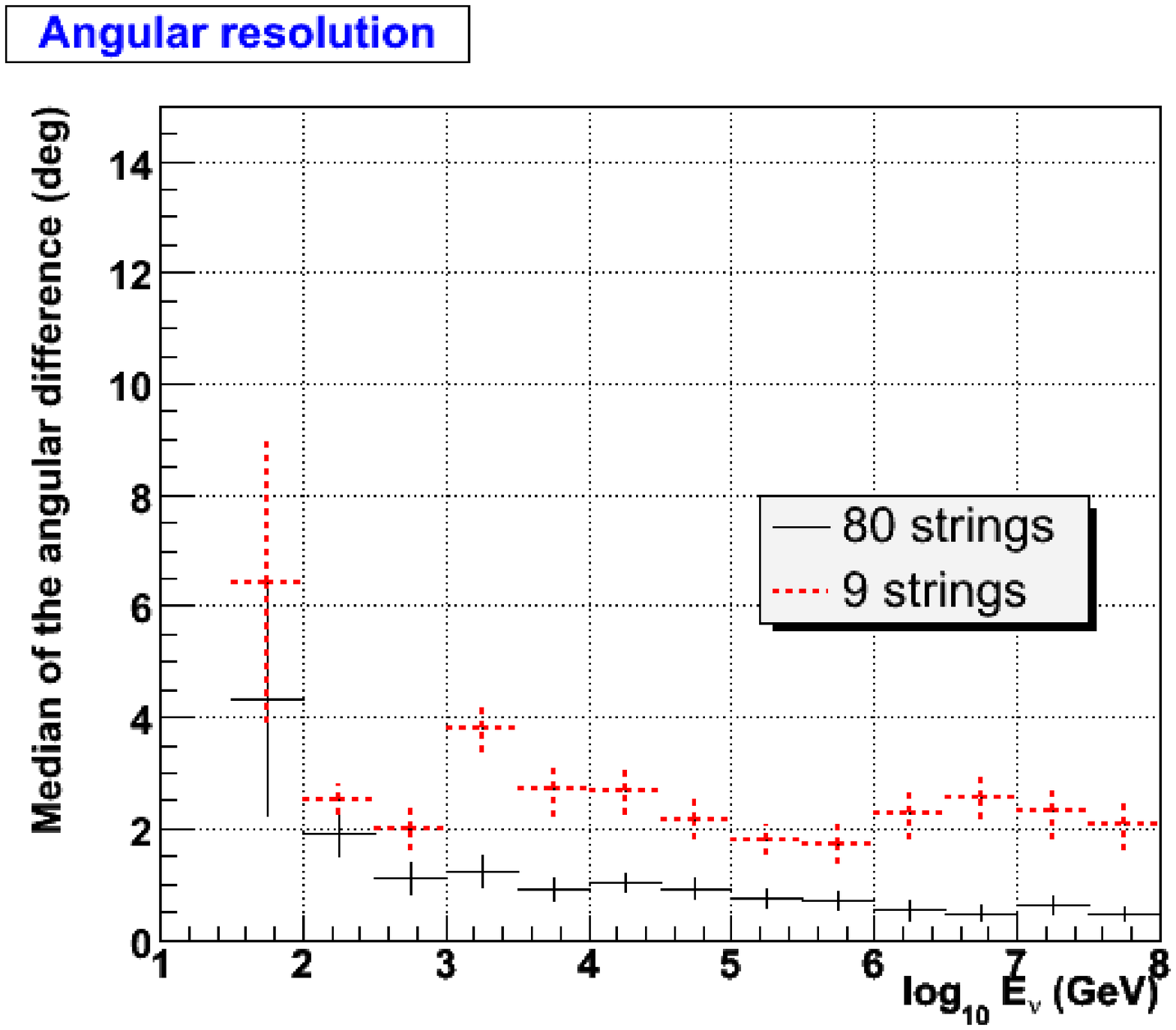}
  \caption{Preliminary study of diffuse muon neutrino effective area and muon track angular resolution for IceCube in the
		present configuration with 9 strings and with the full size. Event selection is the same as from
		\cite{sensitivity}.}
  \label{fig:ic}
\end{figure}

The preliminary event selection has provided 156 observed up-going muons in 90 days livetime (starting from
June 3rd, 2006). This number is consistent with the predicted 144.3 $\pm$ 11.7$_{stat}$ $\pm$ 47.7$_{syst}$ according
to Monte Carlo simulation, of which 137.6 from atmospheric neutrinos, 4.4 from single cosmic muon events and 2.3
from double-uncorrelated cosmic muon events that hit the detector within trigger time window. The preliminary
systematic uncertainty includes contributions from detector simulation approximations, cosmic ray flux and
composition, hadronic interaction models and high energy neutrino cross section. The neutrino effective area
as a function of neutrino energy (see Fig. \ref{fig:ic9}) is the true assessment of detector performance for the
detection of neutrinos. It contains the neutrino interaction probability, muon propagation, detector response and
event selection.

Fig. \ref{fig:ic} shows the neutrino effective area as a function of the neutrino energy (top figure) for the event
selection used in \cite{sensitivity}. The acceptance for 9 strings of IceCube is comparable with AMANDA-II, and the
full IceCube array is expected to increase it by more than one order of magnitude. This increase corresponds to about
a factor of four better angular resolution at high energy (bottom figure). The energy range of the selected events
is about 0.1-5 TeV for atmospheric neutrinos and 1-100 TeV for E$^{-2}$ spectrum.

\subsection{Cosmic Rays}
\label{ss:cr}

Atmospheric neutrinos above $\sim 10^5$ GeV are produced by cosmic rays of energies greater than $\sim 2 - 10 \times 10^5$
GeV, i.e. around the knee \cite{gaisser}. The search for high energy neutrinos from unresolved sources (see \ref{ss:diffuse}) relies
on measuring the excess of neutrinos at high energy with respect to the atmospheric irreducible background. Since the
cosmic ray flux and composition have big uncertainties at and above the knee this is an important component of
systematic uncertainties for the high energy neutrino measurement. The knee is believed to be caused by the escape of lighter mass
cosmic rays from the Galaxy. Due to mass-dependent rigidity cutoff it is expected that the mass composition
would become heavier above the knee. Experimental results, such as the AMANDA/Spase measurement \cite{composition},
seem to confirm this trend. Nevertheless these results seem to be affected by systematic uncertainties that
make interpretation complicated.

\begin{figure}[h]
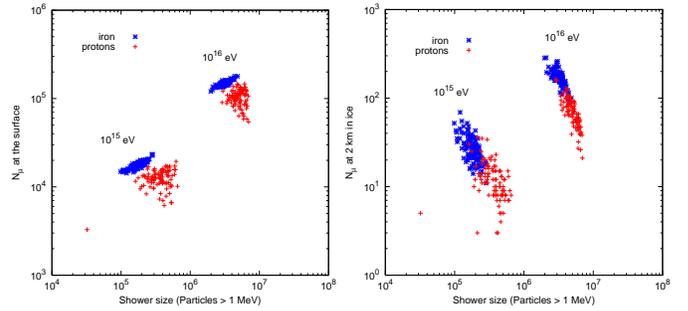

  \centering
  \includegraphics[width=1.7in]{fig/NmuNe-b.epsi}
  \includegraphics[width=1.7in]{fig/NmuNe-a.epsi}
  \caption{Correlation between shower size at the surface and muons.
		Left figure is for muons at the surface.
		Right figure is for muons that reach 2 km depth in the ice ($E_{\mu}~\geq$ 400 GeV). From \cite{gaisser2}}
  \label{fig:cr}
\end{figure}

An efficient way to probe the primary composition is the correlation of muon to electron content in a shower.
The coincident measurement of shower size and muons at the surface provides a better mass resolution (see
Fig \ref{fig:cr} on the left), whereas high energy muons provide a better primary energy resolution (see Fig
\ref{fig:cr} on the right) \cite{engel}.

The IceTop surface array has been designed to measure the cosmic ray flux and composition in the range $\sim 10^5 - 10^9$ GeV.
The digitized waveform from the optical sensors provides an extra degree of freedom to improve the shower core
reconstruction. The estimated acceptance for the present day array geometry is $\sim 0.001$ km$^2$ sr, and it
will be $\sim 0.2$ km$^2$ sr for the full array.

\subsection{Neutrinos from Point Sources}
\label{ss:point}

The search for high energy muon neutrinos from point sources has been performed with AMANDA with different data sets
\cite{97,00,02}. Each of these searches did not reveal any signal, nevertheless the sensitivity has been significantly
improved in time due to geometry size increase, data accumulation, better understanding of detector performance and to
more efficient analysis techniques \cite{ecrs04}. Data collected by AMANDA in the years 2000-04 have been cumulatively
analysed, for a total livetime of 1001 days. This analysis, similarly to the previous ones, has filtered and selected
up-going muons in order to achieve a good background rejection with high retention acceptance for neutrino events,
combined with directional resolution optimization. The 4282 observed up-going events result to be compatible with the predicted
3627 to 4912 atmospheric neutrino-induced events from a full Monte Carlo simulation. The spread of predicted events
accounts for theoretical and experimental systematic uncertainties \cite{04, point}.

The search for point sources has been done using this final up-going muon sample with different methods. For each of them
the expected background has been found using experimental data off-source from the same declination band. %Due to
%the polar position of the array and its azimuthal symmetry, the detection sensitivity depends only on declination.

\begin{figure}[h]
  \centering
  \includegraphics[width=3.2in]{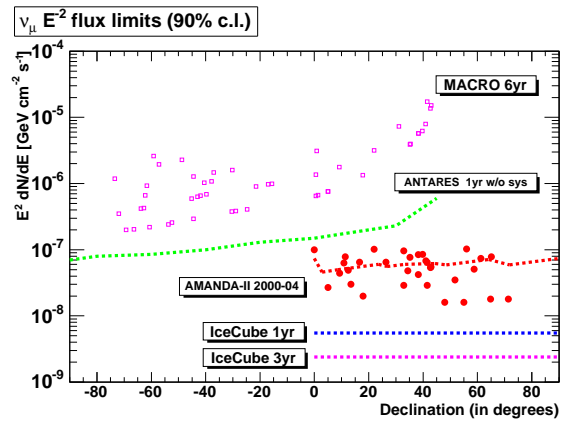}
  \caption{Upper limits on E$^{-2}$ $\nu_{\mu}$ fluxes (90\% CL) vs declination using the AMANDA data collected in 2000-04
		(corresponding to 1001 days livetime)
		for selected sources (full red circles) and sensitivity (red dashed line) \cite{point}. The IceCube estimated
		sensitivity for 1 year is the blue dashed line and for 3 years in the purple dashed line \cite{sensitivity}.
		The MACRO upper limits (open squares)
		are from \cite{macropoint} and the ANTARES preliminary sensitivity for 1 year is from \cite{antares}.}
  \label{fig:point}
\end{figure}

A full-sky search has been performed with a binned method and the 90\% CL sensitivity for a hypothetical E$^{-2}$ flux,
has been determined to be fairly constant in declination and is about E$^2\cdot \Phi < 10^{-7}$ GeV cm$^{-2}$ s$^{-1}$
(see Fig. \ref{fig:point}).
The highest significance observed is 3.7 $\sigma$ and, using scrambled random sky-maps, the probability of seeing something
this significant or higher has been found to be 69\%.

Searches for a set of selected candidates have been also performed, as well as for stacked Active Galactic Nuclei
(AGN) candidates catalogued in different
classes \cite{stacked}. In each case no signal was detected and limits have been calculated for the candidate sources (see
Fig \ref{fig:point}), and for the stacked search limits are about 2-10 better than the one from full sky search.

Specific point source searches have been optimized using correlations with transient phenomena observed in different
electromagnetic wavelengths. Searches of high energy neutrinos in coincidence with Gamma Ray Bursts (GRB) have also been done
\cite{grb}, providing no signal with an excellent acceptance, due to the good background rejection. Also for a source
above the horizon, the Soft Gamma Ray Repeater SGR 1806-20, a search for muon and neutrinos from gamma ray interaction in
the atmosphere has been done \cite{sgr}.

The expected sensitivity for the full IceCube neutrino telescope and for one year is expected to be at least one order
of magnitude better than present 5-year AMANDA sensitivity (see Fig. \ref{fig:point}). This makes IceCube suitable for
probing the most promising galactic and extragalactic source candidates for neutrino emission, especially if combined
in a multi-messanger search campaign, which is under development.

\subsection{Neutrinos from Diffuse Sources}
\label{ss:diffuse}

If individual point sources of neutrinos are too weak even for a km$^3$-scale telescope, other
techniques need to be considered to search for high energy extraterrestrial neutrinos, besides time correlation with external
triggers (for GRBs) and stacking classes of AGNs. We can assume that neutrinos have been emitted by a large number of
unresolved sources at all cosmological times. In this case we might look for a diffuse flux of neutrinos with no space and
time correlation. Under the hypothesis that high energy neutrinos are produced by cosmic ray through
second order Fermi acceleration mechanism, we expect them to have a spectrum like E$^{-2}$, therefore harder than
atmospheric neutrinos ($\propto$ E$^{-3.7}$). The excess of extraterrestrial neutrinos, with respect to the bulk of
detected up-going muon candidates, is expected to occur in the high energy tail. AMANDA has been done diffuse searches
for muon neutrinos and all flavor neutrinos.

\begin{figure}[t]
  \centering
  \includegraphics[width=2.5in,angle=-90]{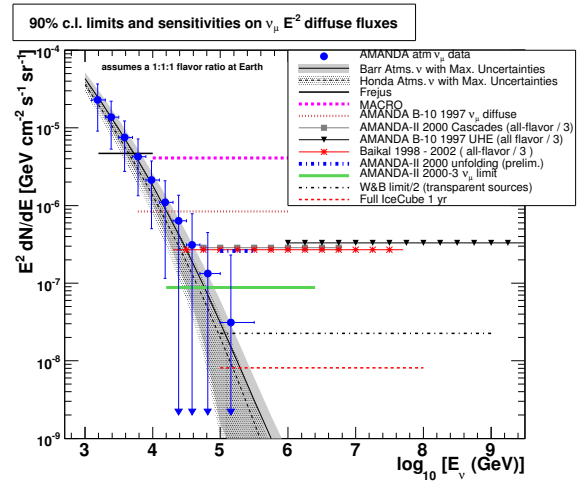}
  \caption{Flux upper limits for diffuse neutrinos with an E$^{-2}$ energy spectrum.
	All-flavor neutrino results are shown along with muon neutrino ones by
	dividing them by three based on the assumption of a 1:1:1 flavor ratio at
	Earth. This figure shows the unfolded energy spectrum of muon neutrinos
	from 2000 AMANDA-II \cite{kirsten}, and the muon neutrino limits from Frejus \cite{frejus},
	MACRO \cite{macrodiff} and AMANDA-B10 \cite{b10}. Also all-flavor neutrino limits
	from AMANDA-II 2000 \cite{a2000}, AMANDA-B10 UHE selection \cite{uhe} and the
	Baikal 5 year limit \cite{baikalim}. The Waxman-Bahcall upper bound \cite{wb}
	and predicted full IceCube sensitivity \cite{performance} are also shown.}
  \label{fig:diffuse}
\end{figure}

The muon neutrino search basically selects a sample of up-going muon tracks compatible with atmospheric neutrinos,
by requesting simply the event to be good quality with good angular resolution. A muon energy estimator is used
to look for the optimum energy cut for selecting E$^{-2}$ neutrinos among the atmospheric ones. Since there is no
off-source background in a diffuse search, Monte Carlo simulated atmospheric and cosmic neutrinos, with event propagation
and detector response, have been used for the energy cut optimization \cite{b10}. Fig. \ref{fig:diffuse} shows the measured
upper limits on E$^{-2}$ fluxes.
Upper limits have been calculated also for neutrino energy spectra other than E$^{-2}$ and the results are being
reported in \cite{diffuse}.

The all-flavor neutrino search selects cascade events in the detector, rather than tracks, and thus it is
sensitive to both charged and neutral current interactions. Searches of contained and semi-contained cascade events
provide a good vertex and energy resolution \cite{a2000}, while searches of non-contained Ultra High Energy (UHE) events
provided extended AMANDA sensitivity up to $\sim 10^9$ GeV \cite{uhe}.

The IceCube neutrino telescope will significantly improve the sensitivity for detecting diffuse high energy neutrinos,
as shown in Fig. \ref{fig:diffuse}, where it appears that Waxman-Bahcall upper bound is in the reach of IceCube.
% sono arrivato qui CONTROLLARE I VALORI

\section{Particle Physics with IceCube}
\label{s:partphys}

The possibility to detect a $\nu_{\tau}$ with a neutrino telescope is intringuing since they are expected to
reach the detector in number comparable to $\nu_e$ and $\nu_{\mu}$ as a consequence of flavor oscillation between their source
and the Earth. In general the flavor ratio of astrophysical neutrinos is expected to be
$\nu_e:\nu_{\mu}:\nu_{\tau} = 1:1:1$, but this ratio might easily be modified by phenomena such as neutron decay,
two-photon annihilation to muon pairs or muon synchrotron cooling inside the source (see \cite{tau} and references
therein).

The measurement of flux of the different neutrino flavors would open a window toward the understanding of particle
physics involved in the neutrino source processes and propagation.

The experimental identification of
a neutrino flavor requires a unique signature that can be discriminated from the background. The identification
of $\nu_{\tau}$ has a clear signature in a neutrino telescope when it undergoes a charged current interaction, and
the produced tau lepton propagates for a given distance before decaying to produce a second cascade inside the detector.
These {\it double bang} events are experimentally detectable in a very narrow energy range, namely $\sim$ 10$^6$-10$^8$
GeV, and require a large volume to be able to reconstruct their unique topology.

IceCube is probing dfferent approaches to increase the sensitivity for detection of $\nu_{\tau}$, such as the
full use of the recorded waveform, through the recognition of the two close cascades. The $\tau$ looses energy at
a higher rate than a muon with the same energy. Therefore investigations are ongoing to understand whether a tau decaying
into a muon can be identified in IceCube.

\section{Conclusion}
The IceCube neutrino telescope has been designed to be a self-calibrating apparatus where each optical sensor is a stand-alone
computer requiring small maintenance while providing high performance for a long operation time. The high dynamic range
of the DOMs, their detailed individual characterization and calibration, low dark noise rate and good timing resolution,
make IceCube suitable for high energy neutrino astronomy. IceCube is expected to significantly improve the sensitivities
achieved by AMANDA, due to its size and the use of full digitized waveforms recorded by the PMTs. Multi-messenger campaigns
are among the main achievements that IceCube is seeking to obtain. The surface array IceTop is designed to extend the
measurement of cosmic ray spectrum and composition up to 10$^9$ GeV, providing an excellent primary energy resolution via
coincident measurement of penetrating muons and shower size at the surface. Finally, starting from 2007, AMANDA becomes
an integrated instrument within the same IceCube Data Acquisition System. The Transient Waveform Records (TWR), i.e. the
digitization at the surface of the PMT pulses recorded by the denser instrumented AMANDA, will provide a good sensitivity
at low energy, not reachable by IceCube only.


\section*{Acknowledgment}% optional entry into table of contents (if used)
%\addcontentsline{toc}{section}{Acknowledgment}
We acknowledge the support from the following agencies: National Science Foundation-Office of Polar Program, National Science Foundation-Physics Division, University of Wisconsin Alumni Research Foundation, Department of Energy, and National Energy Research Scientific Computing Center (supported by the Office of Energy Research of the Department of Energy), the NSF-supported TeraGrid system at the San Diego supercomputer Center (SDSC), and the National Center for supercomputing Applications (NCSA); Swedish Research Council, Swedish Polar Research Secretariat, and Knut and Alice Wallenberg Foundation, Sweden; German Ministry for Education and Research, Deutsche Forschungsgemeinschaft (DFG), Germany; Fund for Scientific Research (FNRS-FWO), Flanders Institute to encourage scientific and technological research in industry (IWT), Belgian Federal Office for Scientific, Technical and Cultural affairs (OSTC); the Netherlands Organisation for Scientific Research (NWO); M. Ribordy acknowledges the support of the SNF (Switzerland); J. D. Zornoza acknowledges the Marie Curie OIF Program (contract 007921).



% http://www.ctan.org/tex-archive/macros/latex/contrib/supported/IEEEtran/bibtex
%\bibliographystyle{IEEEtran.bst}

\newpage
\begin{thebibliography}{1}

\bibitem{baikal}
V.A.~Balkanov et al. (Baikal Collaboration), Astrop. Phys. {\bf 14} (2000) 61

\bibitem{amanda}
E.~Andr\'es, et al. (AMANDA Collaboration), Astrop. Phys. {\bf 13} (2000) 1

\bibitem{gzk}
K.~Greisen, Phys. Rev. Lett {\bf 16} (1966) 748-750\\
G.T.~Zatsepin and V.A.~Kuz'min, J. Exp. Theor. Phys. Lett {\bf 4} (1966) 78-80 [ZhETF Pis'ma {\bf 4} (1966) 114-117]

\bibitem{hess}
J.~Hinton, et al. (H.E.S.S. Collaboration), arXiv:astro-ph/0607351

\bibitem{bere}
V.S.~Berezinsky and V.I.~Dokuchaev, arXiv:astro-ph/0002274

\bibitem{panic05}
P.~Desiati for the IceCube Collaboration, {\it in Proc. of PANIC05, Santa Fe (NM), USA, AIP Conference Proceedings,
Edited by P.D. Barnes et al.} (2006) 983

\bibitem{hill}
G.C.~Hill for the IceCube Collaboration, {\it to appear in Proc. of Neutrino 2006, Santa Fe (NM), USA} (2006)

\bibitem{wb}
E.~Waxman and J.N.~Bahcall, Astrophys. J. {\bf 541} (2000) 707

\bibitem{performance}
A.~Achterberg, et al. (IceCube Collaboration), Astrop. Phys. {\bf 26} (2006) 3, 155-230

\bibitem{karle}
A.~Karle for the IceCube Collaboration, {\it to appear in Proc. of VLVnT2, Oct. 2005, Catania, Italy} (2006),
arXiv:astro-ph/0608139

\bibitem{opto}
E.~Andr\'es, et al. (AMANDA Collaboration), J. Geophys. Res. {\bf 111} (2006) D13203

\bibitem{logger}
N.E.~Bramall et al., Geophys. Res. Lett., {\bf 32} (2005) L21815

\bibitem{barr}
G.D.~Barr et al., Phys. Rev. D {\bf 70} (2004) 023006

\bibitem{honda}
M.~Honda et al., Phys. Rev. D {\bf 70} (2004) 043008

\bibitem{kirsten}
K.~M\"unich for the IceCube Collaboration, {\it in Proc. of 29th ICRC, Pune, India} (2005)

\bibitem{gaisser}
T.K.~Gaisser, Astrop. Phys. {\bf 16} (2002) 285-294, arXiv:astro-ph/0104327

\bibitem{engel}
R.~Engel, {\it in Proc. of Workshop on Physics at the End of the Galactic Cosmic Ray Spectrum} (2005)

\bibitem{composition}
J.~Ahrens et al. (AMANDA and SPASE Collaborations), Astrop. Phys. {\bf 21} (2005) 565-581

\bibitem{gaisser2}
T.K.~Gaisser, {\it to appear in Proc. of International Workshop on
Energy Budget in the High Energy Universe, Chiba, Japan} (2006)

\bibitem{97}
J.~Ahrens et al. (AMANDA Collaboration), Astrophys. J. {\bf 583} (2003) 1040

\bibitem{00}
J.~Ahrens et al. (AMANDA Collaboration), Phys. Rev. Lett. {\bf 92} (2004) 071102

\bibitem{02}
M.~Ackermann et al. (AMANDA Collaboration), Phys. Rev. D {\bf 71} (2005) 077102

\bibitem{ecrs04}
P.~Desiati for the IceCube Collaboration, {\it in Proc. of ECRS04, Florence, Italy, Intern. Journ. of Mod. Phyd. A,
Edited by  O. Adriani et al.} (2005) 6919

\bibitem{04}
M.~Ackermann for the IceCube Collaboration, {\it to appear in Proc. of The Multi-messenger approach to
high-energy gamma ray sources, Barcellona, Spain} (2006)

\bibitem{point}
A.~Achterberg et al., {\it Search for point sources of astrophysical neutrinos with the AMANDA neutrino telescope},
in preparation

\bibitem{sensitivity}
J.~Ahrens et al. (IceCube Collaboration), Astrop. Phys. {\bf 20} (2004) 507

\bibitem{macropoint}
M.~Ambrosio et al. (MACRO Collaboration), Astrophys. J. {\bf 546} (2001) 1038

\bibitem{antares}
T.~Montaruli et al. for ANTARES Collaboration, Acta Phys. Polon. {\bf B36} (2005) 509, arXiv:hep-ph/0410079

\bibitem{stacked}
A.~Achterberg et al. for IceCube Collaboration, Astrop. Phys., {\bf 26} 4-5 (2006) 282-300

\bibitem{grb}
M.~Stamatikos et al. for IceCube Collaboration, {\it in Proc. of 29th International Cosmic Ray Conference in Pune, India} (2005), arXiv:astro-ph/0510336\\
B.~Hughey, I.~Taboada et al., {\it in Proc. of 29th International Cosmic Ray Conference in Pune, India} (2005), astro-ph/0509570

\bibitem{sgr}
A.~Achterberg et al. (IceCube Collaboration), Astrop. Phys. accepted

\bibitem{diffuse}
A.~Achterberg et al. (IceCube Collaboration), in preparation

\bibitem{frejus}
W.~Rhode et al., Astrop. Phys. {\bf 4} (1996) 217

\bibitem{macrodiff}
M.~Ambrosio et al. (MACRO Collaboration), Astrop. Phys. {\bf 19} (2003) 1

\bibitem{b10}
J.~Ahrens et al. (AMANDA Collaboration), Phys. Rev. Lett. {\bf 90} (2003) 251101

\bibitem{a2000}
M.~Ackermann et al. (AMANDA Collaboration), Astrop. Phys. {\bf 22} (2004) 127

\bibitem{uhe}
M.~Ackermann et al. (AMANDA Collaboration), Astrop. Phys. {\bf 22} (2004) 339

\bibitem{baikalim}
V.~Aynutdinov et al. (Baikal Collaboration), Astrop. Phys. {\bf 25} (2006) 140

\bibitem{tau}
T.~DeYoung, S.~Razzaque and D.F.~Cowen, arXiv:astro-ph/0608486



%\bibitem{halzen1}
%F.~Halzen, \emph{Astroparticle Physics with High Energy Neutrinos: from AMANDA to IceCube}, arXiv:astro-ph/0602132
%3rd~ed.\hskip 1em plus 0.5em minus 0.4em\relax Harlow, England: Addison-Wesley, 1999.

\end{thebibliography}

\end{document}